\documentclass{pasj00}
\draft

\begin{document}
\SetRunningHead{M. Sakai et al.}{X-ray observation of PSR~J1648$-$4611}
%\Received{}%{yyyy/mm/dd}
%\Accepted{}%{yyyy/mm/dd}
%\Published{}%{yyyy/mm/dd}

\title{Discovery of Diffuse Hard X-Ray Emission from the Vicinity of PSR~J1648$-$4611 with Suzaku}

 \author{
   Michito \textsc{Sakai},\altaffilmark{1}
   Hironori \textsc{Matsumoto},\altaffilmark{2}
   Yoshito \textsc{Haba},\altaffilmark{2}
   Yasufumi \textsc{Kanou},\altaffilmark{1}
   and
   Youhei \textsc{Miyamoto}\altaffilmark{1}}
\altaffiltext{1}{Division of Particle and Astrophysical Science, Graduate School of Science, Nagoya University, \\Furo-cho, Chikusa-ku, Nagoya 464-8602}
\email{m\_sakai@u.phys.nagoya-u.ac.jp}
\altaffiltext{2}{Kobayashi-Maskawa Institute for the Origin of Particles and the Universe, Nagoya University, \\Furo-cho, Chikusa-ku, Nagoya 464-8602}


\maketitle

\begin{abstract}
We observed the pulsar PSR~J1648$-$4611 with Suzaku.  Two
X-ray sources, Suzaku~J1648$-$4610 (Src~A) and
Suzaku~J1648$-$4615 (Src~B), were found in the field of
view.  Src~A is coincident with the pulsar PSR~J1648$-$4611,
which was also detected by the Fermi Gamma-ray Space
Telescope.  A hard-band image indicates that Src~A is
spatially extended. We found point sources in the vicinity
of Src~A by using a Chandra image of the same region, but
the point sources have soft X-ray emission and cannot
explain the hard X-ray emission of Src~A. The hard-band
spectrum of Src~A can be reproduced by a power-law model
with a photon index of $2.0^{+0.9}_{-0.7}$.  The X-ray flux
in the 2--10 keV band is $1.4 \times
10^{-13}$~erg~cm$^{-2}$~s$^{-1}$.  The diffuse emission
suggests a pulsar wind nebula around PSR~J1648$-$4611, but
the luminosity of Src~A is much larger than that expected
from the spin-down luminosity of the pulsar.  Parts of the
very-high-energy $\gamma$-ray emission of HESS~J1646$-$458
may be powered by this pulsar wind nebula driven by
PSR~J1648$-$4611. Src~B has soft emission, and its X-ray
spectrum can be described by a power-law model with a photon
index of $3.0^{+1.4}_{-0.8}$. The X-ray flux in the
0.4--10~keV band is $6.4 \times
10^{-14}$~erg~s$^{-1}$~cm$^{-2}$. No counterpart for Src~B 
is found in literatures.
\end{abstract}

\section{Introduction}

The HESS (High Energy Stereoscopic System) Collaboration has
recently reported the detection of the degree-scale extended
very-high-energy (VHE) $\gamma$-ray source
HESS~J1646$-$458~\citep{Abr12}.  The VHE $\gamma$-ray source
is centered on the massive stellar cluster Westerlund
1~\citep{Wes61}.  \citet{Abr12} proposed two scenarios.  One
is a single-source scenario, where Westerlund 1 is favored
as site of VHE particle acceleration.  Here, a hadronic
parent population would be accelerated within the stellar
cluster.  The other is a multi-source origin, where a
scenario involving the pulsar PSR~J1648$-$4611~\citep{Kra03}
could be viable to explain parts of the VHE $\gamma$-ray
emission of HESS~J1646$-$458.

PSR~J1648$-$4611 is a radio pulsar discovered by the Parkes Multibeam
Pulsar Survey~\citep{Kra03}.
It is located at the position
($l$, $b$) =
(\timeform{339.4383D}, \timeform{-00.7938D})
\footnote{($\alpha$, $\delta$)$_{\rm J2000.0}$
$=$(\timeform{16h48m22.0s}, $-$\timeform{46D11'16.0"}).}
with a pulse period of $P = 0.1649$ s and a period derivative of
$\dot P = 23.7 \times 10^{-15}$ {\rm s s$^{-1}$}.
The distance to the source was estimated to be $d = 5.7$ kpc from
the pulsar's dispersion measure using the \citet{Tay93} model
for the Galactic distribution of free electrons.
The characteristic age and the spin-down luminosity are
$\tau_{c} = 1.1 \times 10^{5}$ yr and
$\dot E = 2.1 \times 10^{35}$ {\rm erg s$^{-1}$}, respectively.

Recently, the Large Area Telescope (LAT) on the Fermi
Gamma-ray Space Telescope (Fermi) detected $\gamma$-ray
source that is spatially associated with the
PSR~J1648$-$4611 (2FGL~J1648.4$-$4612; \cite{Nol12}).
A $\gamma$-ray pulse detection from this pulsar is also
reported in the Public List of LAT-Detected Gamma-Ray Pulsars.\footnote{See
  $\langle$https:\slash\slash{}confluence.slac.stanford.edu\slash{}display\slash{}GLAMCOG\slash{}Public+List+of+LAT-Detected+Gamma-Ray+Pulsars$\rangle$.}
These results imply that the origin of the VHE $\gamma$-ray
emission also relates to this pulsar.  Active
pulsars are losing a significant part of energy via
relativistic particles, and form pulsar wind nebulae (PWNe).
PWNe emit synchrotron emission from the radio to X-ray
bands.  In addition, some PWNe are known to be VHE
$\gamma$-ray emitters.  Thus, PSR~J1648$-$4611 is a possible
candidate for the source of the energy of HESS~J1646$-$458.
However, X-ray emission from PSR~J1648$-$4611 has not been
reported.

In the following sections, the results of the Suzaku
observation of PSR~J1648$-$4611 will be discussed; in
addition, our own analysis on the Chandra observation of
this region will be utilized to estimate the contribution of
point sources.  In this paper, uncertainties are described
at the 90\% confidence level, while the errors of data
points in X-ray spectra and radial profiles are at the 1$\sigma$ level, 
unless otherwise stated.

\section{Observations and Data Reduction}

\subsection{Suzaku}

We observed the PSR~J1648$-$4611 region with Suzaku~\citep{Mit07}
on 2010 September 23 and 24 (OBSID=505051010).
The observation was performed with the three CCD cameras (XISs:~\cite{Koy07})
located in the focal planes of the X-ray telescopes (XRTs:~\cite{Ser07})
and the non-imaging detector (HXD:~\cite{Kok07,Tak07}).
One of the XIS sensors (XIS 1) has a back-illuminated (BI) CCD, while
the other two (XISs 0 and 3) utilize front-illuminated (FI) CCDs.
Because one of the FIs (XIS 2) suffered catastrophic damage on 2006
November 9, no useful data were obtained.
The XIS was operated in the normal clocking mode (without the Burst or
Window options) with Spaced-raw Charge Injection (SCI)~\citep{Uch09}.

We analyzed the data with the processing version of
2.5.16.28,\footnote{See
  $\langle$http:\slash\slash{}www.astro.isas.jaxa.jp\slash{}suzaku\slash{}process\slash{}history\slash{}v251628.html$\rangle$.}
utilizing the HEADAS software (version 6.11.1) and
calibration database (CALDB) released on 2011 November 10.
All data affected by the South Atlantic Anomaly and
telemetry saturation were excluded.  We excluded the data
obtained with an elevation angle from the Earth rim of $<
5^{\circ}$.  Additionally, for the XIS data, we also
excluded the data obtained with that from the bright Earth
rim of $< 20^{\circ}$ and removed hot/flickering pixels.
After these data screenings, the effective exposure for the
XIS data was 50.2 ks, and that for the HXD data was 40.8 ks.
In this paper, we concentrate on the XIS data.

\subsection{Chandra}

Chandra observed the PSR~J1648$-$4611 region on 2010 January 24 (OBSID=11836).
The exposure time was 10.2~ks.
Chandra has a superior angular resolving capability.
We analyzed Chandra data in order to estimate the contribution of point
sources and the intensity upper limit of an undetected source,
using the Chandra Interactive Analysis of Observations (CIAO) software
version 4.3.1 with CALDB version 4.4.6.

\section{Analysis and Results}

\subsection{Images}

We extracted XIS images in the soft- and hard-energy bands
from each sensor using the screened data.  For the FI
sensors, the soft- and hard-bands are defined as 0.4--3 keV
and 3--10 keV, respectively, while those for the BI sensor
are defined as 0.3--3 keV and 3--7 keV, respectively.  The
corners of the CCD chips illuminated by the $^{55}$\rm{Fe}
calibration sources were excluded.  Non-X-ray backgrounds
(NXB) generated with {\tt xisnxbgen}~\citep{Taw08} were
subtracted from the images.  Then, the soft- and hard-band
images were divided by flat sky images simulated at 1.49 and
4.5 keV using the XRT+XIS simulator {\tt
  xissim}~\citep{Ish07} for vignetting corrections.  The
images were binned by a factor of 8. The images from the two
FI sensors were summed.

\begin{figure*}
  \begin{tabular}{cc}
    \begin{minipage}{0.5\hsize}
      \begin{center}
        \FigureFile(80mm,80mm){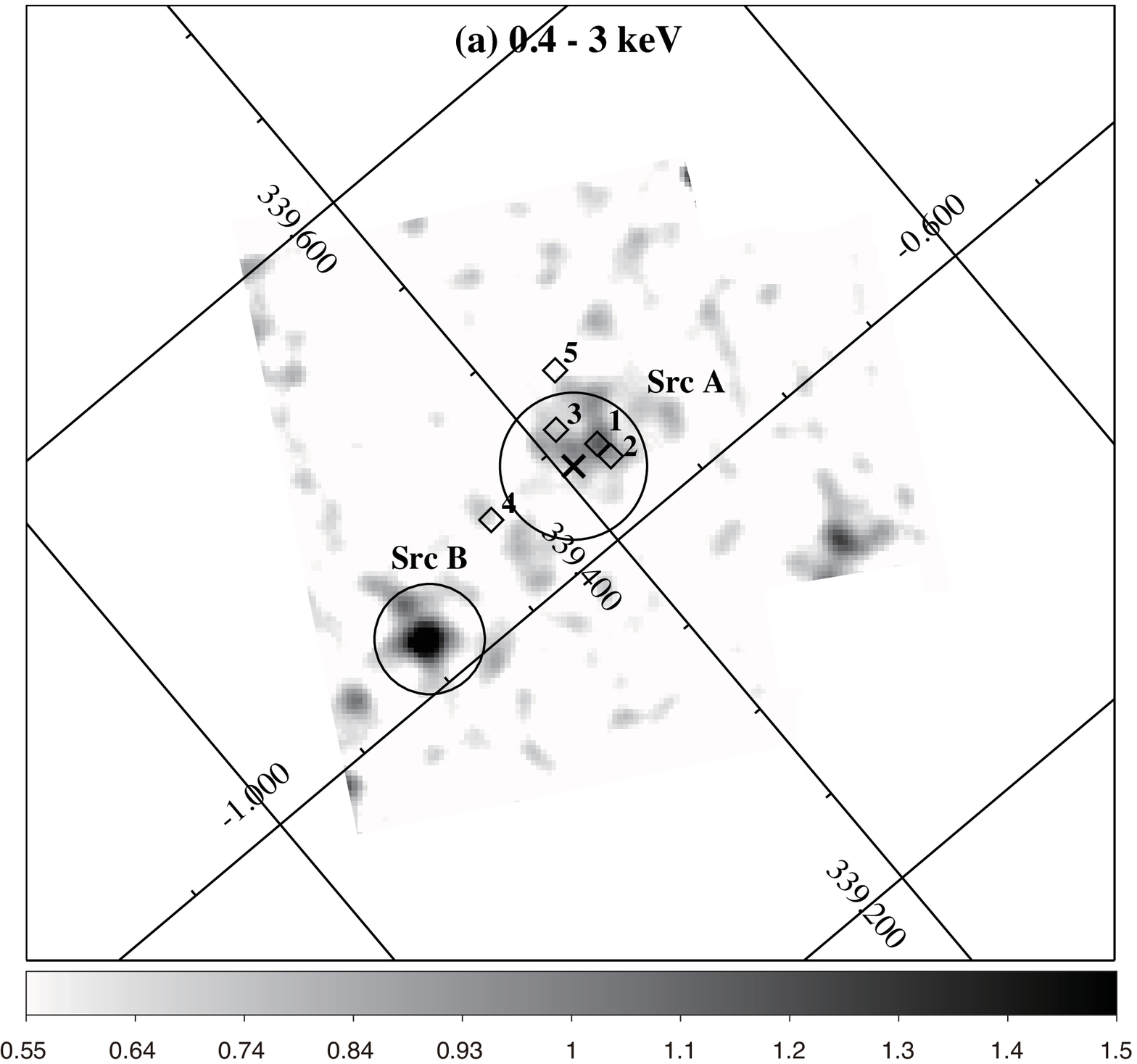}
      \end{center}
    \end{minipage}
    \begin{minipage}{0.5\hsize}
      \begin{center}
        \FigureFile(80mm,80mm){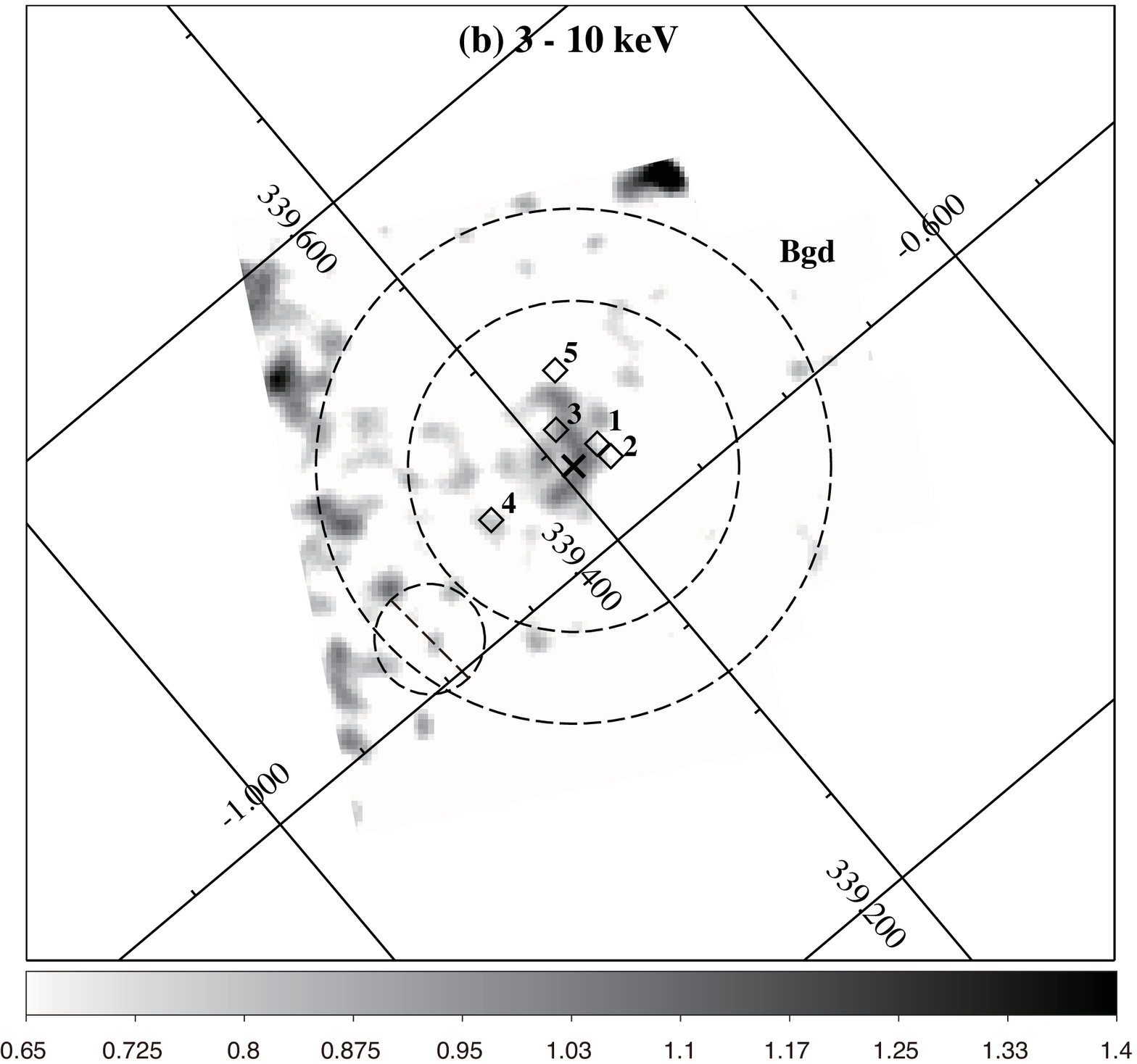}
      \end{center}
    \end{minipage}
  \end{tabular}
  \caption{
Suzaku XIS FI (XIS 0+3) images of the PSR~J1648$-$4611
region in the Galactic coordinates: (a) 0.4--3 keV and (b)
3--10 keV bands. The images were smoothed using a Gaussian
function with $\sigma$ = \timeform{0'.28}. A vignetting
correction was applied after subtracting the NXB, as
described in the text. The cross mark represents the
position of PSR~J1648$-$4611. The numbered diamond points
mark the Chandra point sources, whose properties are listed
in table \ref{tab:1st}. The source regions for the spectrum
of Src~A and Src~B are shown by the solid circles. The
background region is shown by the dashed annulus excluding
the source region of Src~B.
  }\label{fig:1st}
\end{figure*}

Figure \ref{fig:1st} shows the XIS FI images of the
PSR~J1648$-$4611 region smoothed using a Gaussian function
with $\sigma$ = \timeform{0'.28}.  The XIS BI images were
essentially the same, except for the poorer statistics.  
In the hard-band image, 
an X-ray source with a peak position of
($l$, $b$) = (\timeform{339.44D}, \timeform{-0.79D})
\footnote{($\alpha$, $\delta$)$_{\rm J2000.0}$
= (\timeform{16h48m20s}, \timeform{-46D10'57"}).}
was found, and was designated as Suzaku J1648$-$4610 (Src~A).
The peak position of Src~A is close to PSR~J1648$-$4611.
The spatial distribution of Src~A in the soft-band image 
seems to be different from that in the hard-band image
(Figures \ref{fig:1st}(a) and (b)).
The peak position in the soft-band image is
($l$, $b$) = (\timeform{339.44D}, \timeform{-0.78D})
\footnote{($\alpha$, $\delta$)$_{\rm J2000.0}$
= (\timeform{16h48m18s}, $-$\timeform{46D10'32"}).},
and is close to Src~1 found in a Chandra image
(see the end of section 3.1).

Another X-ray source, found at 
($l$, $b$) =
(\timeform{339.42D}, \timeform{-0.90D})
\footnote{($\alpha$, $\delta$)$_{\rm J2000.0}$
= (\timeform{16h48m45s}, \timeform{-46D15'57"}).},
was also bright in the soft X-ray band.
We designated this source as Suzaku J1648$-$4615 (Src~B).
Src~B was not conspicuous in the hard-band image.

\begin{figure}
  \begin{center}
    \FigureFile(160mm,){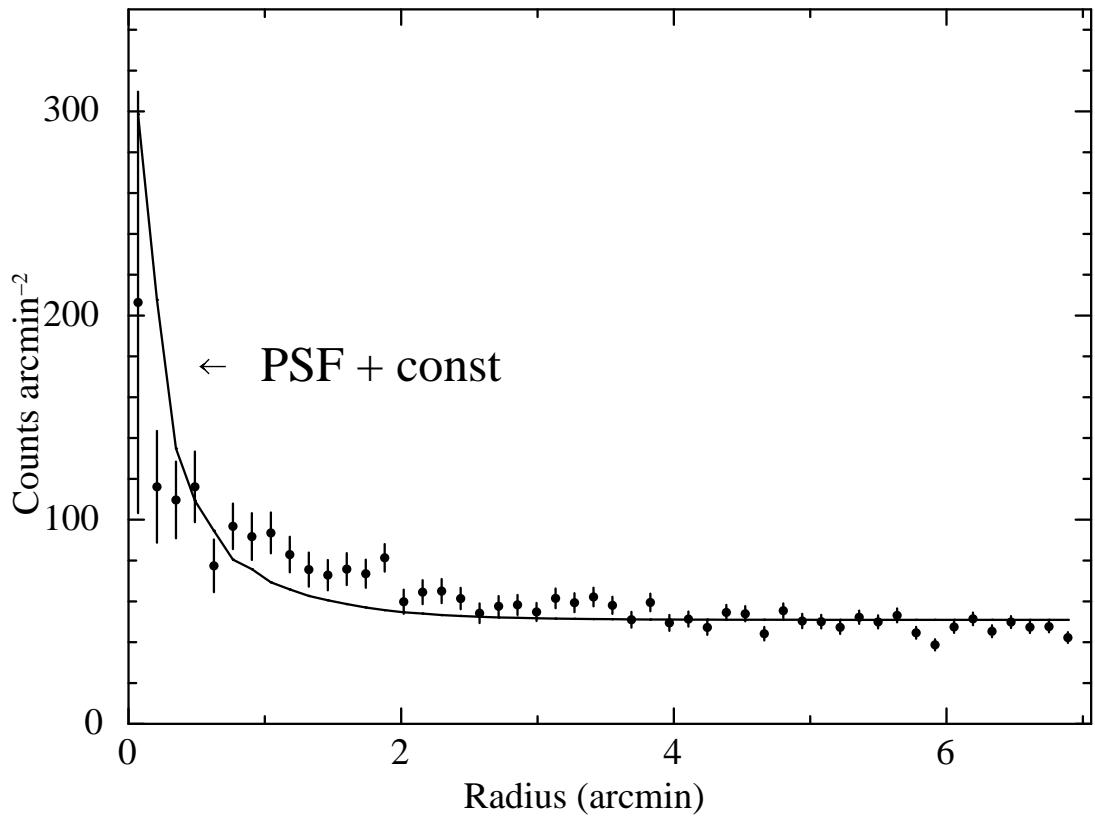}
%    \FigureFile(80mm,80mm){figure2.eps}
  \end{center}
  \caption{
Radial profile of Src~A extracted from the 3--10 keV band
image of the XIS FI sensor (XIS 0+3). The solid line
represents the XIS PSF profile with a constant component.
}\label{fig:2nd}
\end{figure}

We created a radial profile of Src~A in the 3--10 keV band
as shown in Figure~\ref{fig:2nd}, and the profile was
compared with a point-spread function (PSF).  The origin of
the radial profile was the peak in the hard-band image.  As
for the PSF, we obtained the radial profile using the SS Cyg
data observed on 2005 November 2 (OBSID=400006010), which
are the verification phase data for the imaging capability
of the XRT~\citep{Ser07}.  Since the energy dependence of
the PSF is negligible~\citep{Ser07}, the radial profile was
extracted from the 0.4--10 keV band.  In this analysis, the
NXB subtraction and vignetting correction were not applied
to both the radial profiles of Src~A and the PSF.  As shown
in Figure~\ref{fig:2nd}, the profile cannot be fitted
adequately with the PSF plus a constant component model
($\chi^2/d.o.f.$ = 146.1/48), and therefore Src~A must be
diffuse emission or unresolved multiple sources.

\begin{figure}
  \begin{center}
    \FigureFile(160mm,){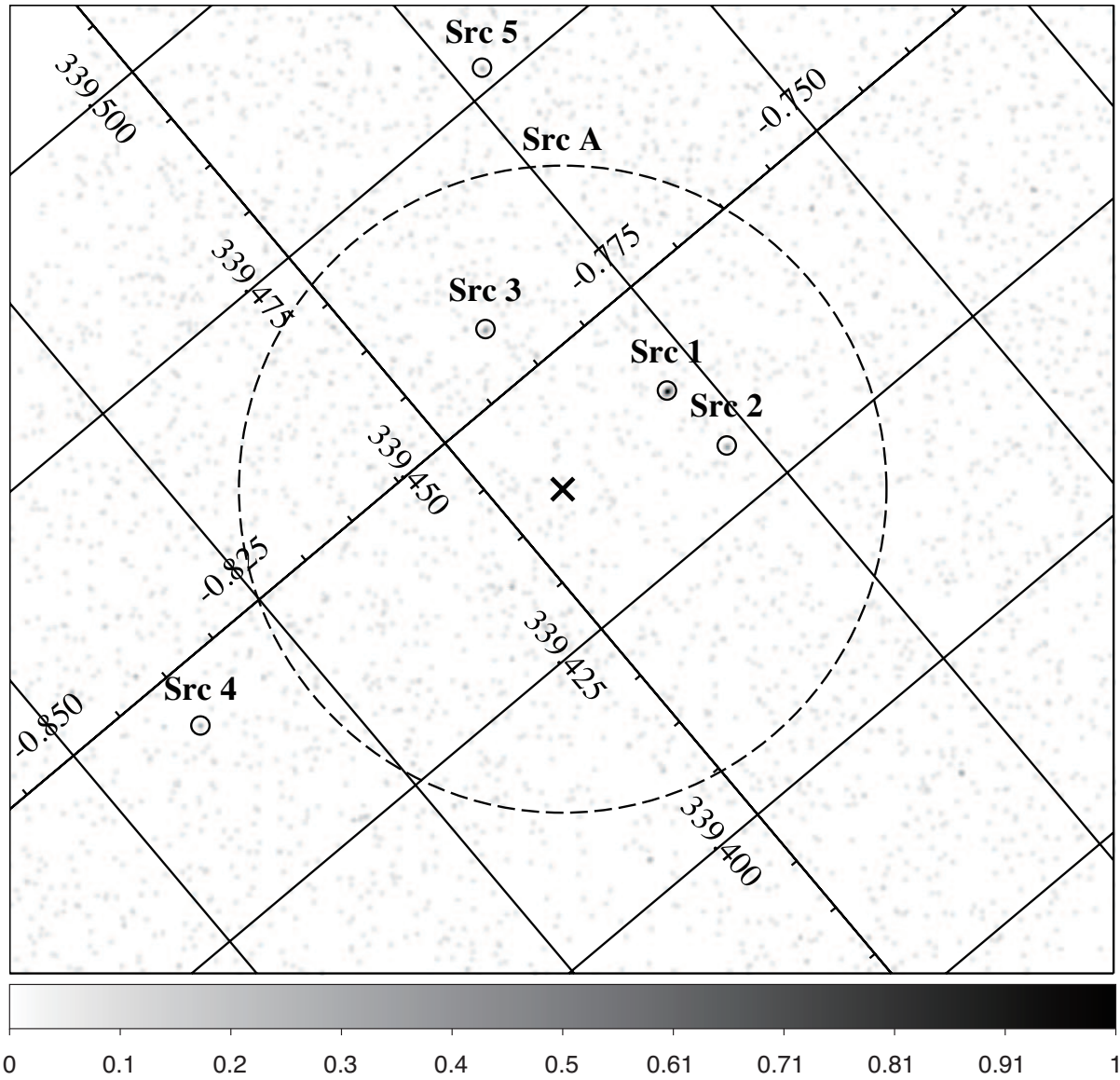}
%    \FigureFile(80mm,80mm){figure3.eps}
  \end{center}
  \caption{
Chandra image around PSR~J1648$-$4611 in the 0.4--10
keV band including the XIS source region for Src~A (dashed circle).
Point sources are indicated by solid circles with
the source numbers. Src~1 is coincident with HD 151228. The
cross mark represents the position of
PSR~J1648$-$4611. Lines of constant Galactic latitude and
longitude are plotted and labeled in the interior of the
figure.
  }\label{fig:3rd}
\end{figure}

Figure \ref{fig:3rd} shows the Chandra image around Src~A in
the 0.4--10 keV band. Several sources can be seen around
Src~A.  A point source search was carried out in the image
with {\tt wavdetect}.  This tool uses a wavelet
method~\citep{Fre02}.  The wavelet scales of {\tt wavdetect}
were the $\sqrt{2}$ series: 1.0 1.414 2.0 2.828 4.0 5.657
8.0 11.314 16.0.  The threshold significance was set to
$10^{-6}$, this is equivalent to stating that the expected
number of false sources per the ACIS-I3 CCD chip is one.  We resolved 5
point sources from the whole region of the ACIS-I3 CCD chip.
Their positions and counts are listed in table
\ref{tab:1st}.  The detected point sources around Src~A are
indicated by solid circles with the source numbers in figure
\ref{fig:3rd}.  From the SIMBAD Astronomical Database
operated at CDS, Strasbourg, France, we found a counterpart
of Src~1 at ($l$, $b$) = (\timeform{339.4396D},
\timeform{-00.7792D})
\footnote{($\alpha$, $\delta$)$_{\rm J2000.0}$
$=$(\timeform{16h48m18.4s}, $-$\timeform{46D10'38.5"}).}, 
which is an A0 Star: HD 151228.

\begin{table*}[htbp]
  \begin{center}
    \caption{Chandra point sources around the field of PSR~J1648$-$4611.}\label{tab:1st}
    \begin{tabular}{lccc} \hline
      Source numbers  & $RA$ (J2000.0) & $Dec$ (J2000.0) & Counts\footnotemark[$*$] (0.4 -- 10 keV) \\ \hline
      Src~1 & 16 48 18.30 & -46 10 39.4 & 26 $\pm$ 5 \\
      Src~2 & 16 48 16.16 & -46 10 59.7 & 7 $\pm$ 3 \\
      Src~3 & 16 48 24.78 & -46 10 16.6 & 8 $\pm$ 3 \\
      Src~4 & 16 48 34.95 & -46 12 43.7 & 4 $\pm$ 2 \\
      Src~5 & 16 48 24.91 & -46 08 39.6 & 3 $\pm$ 2 \\ \hline
      \multicolumn{4}{@{}l@{}}{\hbox to 0pt{\parbox{180mm}{\footnotesize 
            \footnotemark[$*$] The sum of all counts in the source cell minus the sum of the estimated background counts.
          }\hss}}
    \end{tabular}
  \end{center}
\end{table*}

\subsection{Spectra}

\subsubsection{Src~A}

\begin{figure}
  \begin{center}
    \FigureFile(160mm,){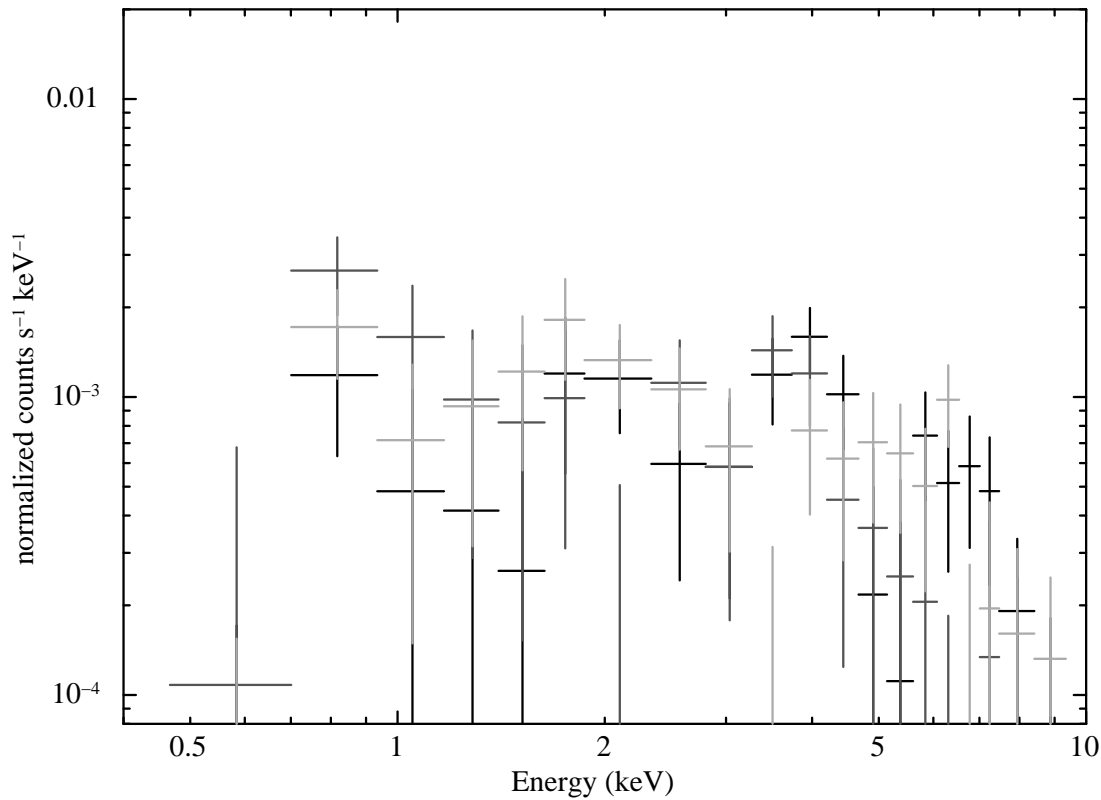}
%    \FigureFile(80mm,80mm){figure4.eps}
  \end{center}
  \caption{XIS spectra of Src~A. Black, Dark Gray and Light Gray lines represent the data for the XIS0, XIS1 and XIS3, respectively.}\label{fig:4th}
\end{figure}

An X-ray spectrum of Src~A of each XIS sensor was extracted
from a circular region with a radius of \timeform{2'}.  An
annulus region with an inner radius of \timeform{4.5'} and
an outer radius of \timeform{7'} was used for a background
spectrum.  NXB spectra in the source and background regions
were constructed using {\tt xisnxbgen}~\citep{Taw08}, and
the NXB were subtracted from the source and background
spectra.  When the background spectrum was subtracted from
the source spectrum, a vignetting correction was taken into
account. The spectra of Src~A thus obtained are shown in
figure \ref{fig:4th}. In the spectrum analysis below,
Redistribution Matrix Files (RMFs) and Ancillary Response
Files (ARFs), made using the softwares {\tt xisrmfgen} and
{\tt xissimarfgen}~\citep{Ish07}, were used.

\begin{figure}
  \begin{center}
    \FigureFile(160mm,){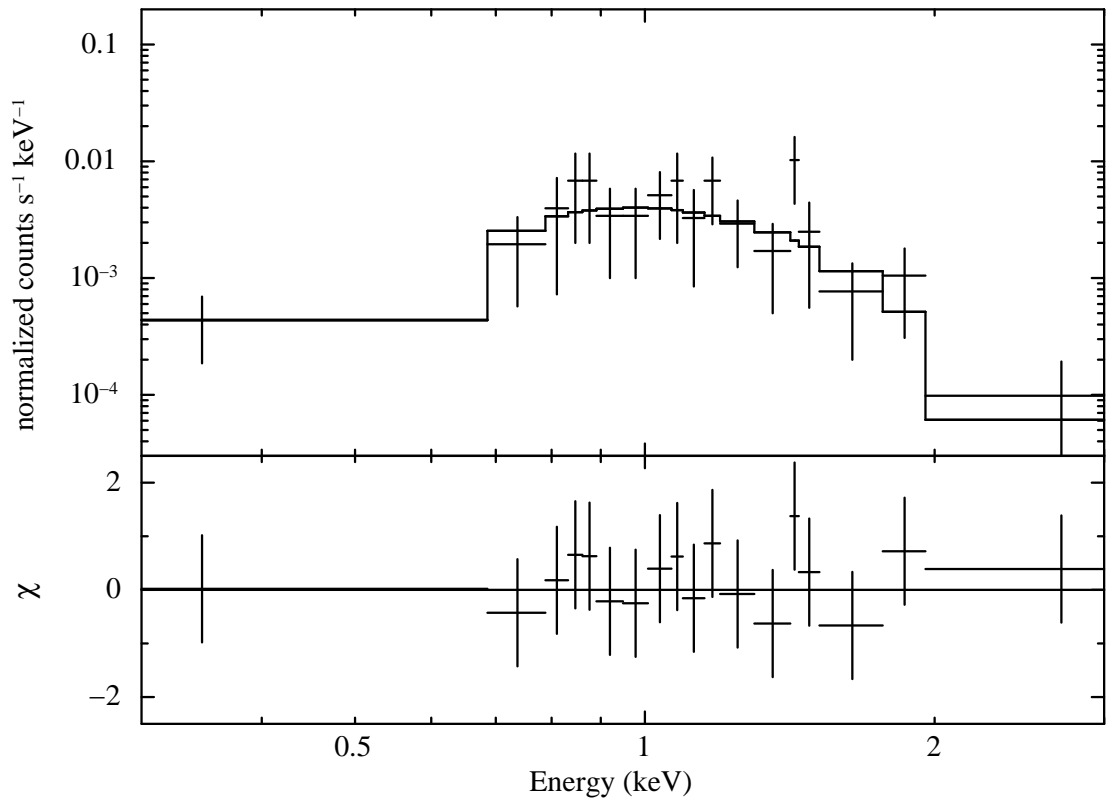}
%    \FigureFile(80mm,80mm){figure5.eps}
  \end{center}
  \caption{Combined Chandra spectrum of three point sources
(Srcs 1, 2 and 3). Black lines represent the best-fit model.}\label{fig:5th}
\end{figure}

The Chandra image (figure \ref{fig:3rd}) shows that the
Src~A region includes three point sources.  A Chandra spectrum
including all three sources was constructed in order to
estimate the contribution of the point sources to the Suzaku
spectra.  Circular regions with a radius of \timeform{3.4"}
centered on each source were used for the source spectrum,
and annular regions with an inner radius of \timeform{5"}
and an outer radius of \timeform{12.6"} around each source
were used for a background spectrum.  The spectrum thus
obtained (figure \ref{fig:5th}) is soft where almost no
X-ray photon can be seen above 2 keV.  Note that the Suzaku
soft-band image seems to follow the distribution of the
Chandra point sources, while the hard-band image shows a
different distribution (figure~\ref{fig:1st}).  This suggests that the soft
emission of Src~A is attributed to the Chandra sources and
that the origin of the hard emission of Src~A is different
from that of the soft emission.  We fitted a bremsstrahlung
model with interstellar absorption to the Chandra spectrum
of point sources in the 0.4--2 keV band.  The cross sections
of the photoelectric absorption were obtained from
\citet{Mor83}.  A hydrogen-equivalent column density $N_{\rm
  H}$, plasma temperature $kT$, and a normalization were set
to be free parameters.  The best-fit parameters are $N_{\rm
  H} = 0.1^{+0.7}_{-0.1} \times 10^{22}$ {\rm cm}$^{-2}$ and
$kT = 0.45^{+0.60}_{-0.28}$ keV, and $\chi2/d.o.f.$=5.99/15.  The observed flux in the
0.4--2 keV band is $F$(0.4--2 keV) = $(3.4 \pm 1.0) \times
10^{-14}$ {\rm erg cm$^{-2}$ s$^{-1}$}.

\begin{figure}
  \begin{center}
    \FigureFile(160mm,){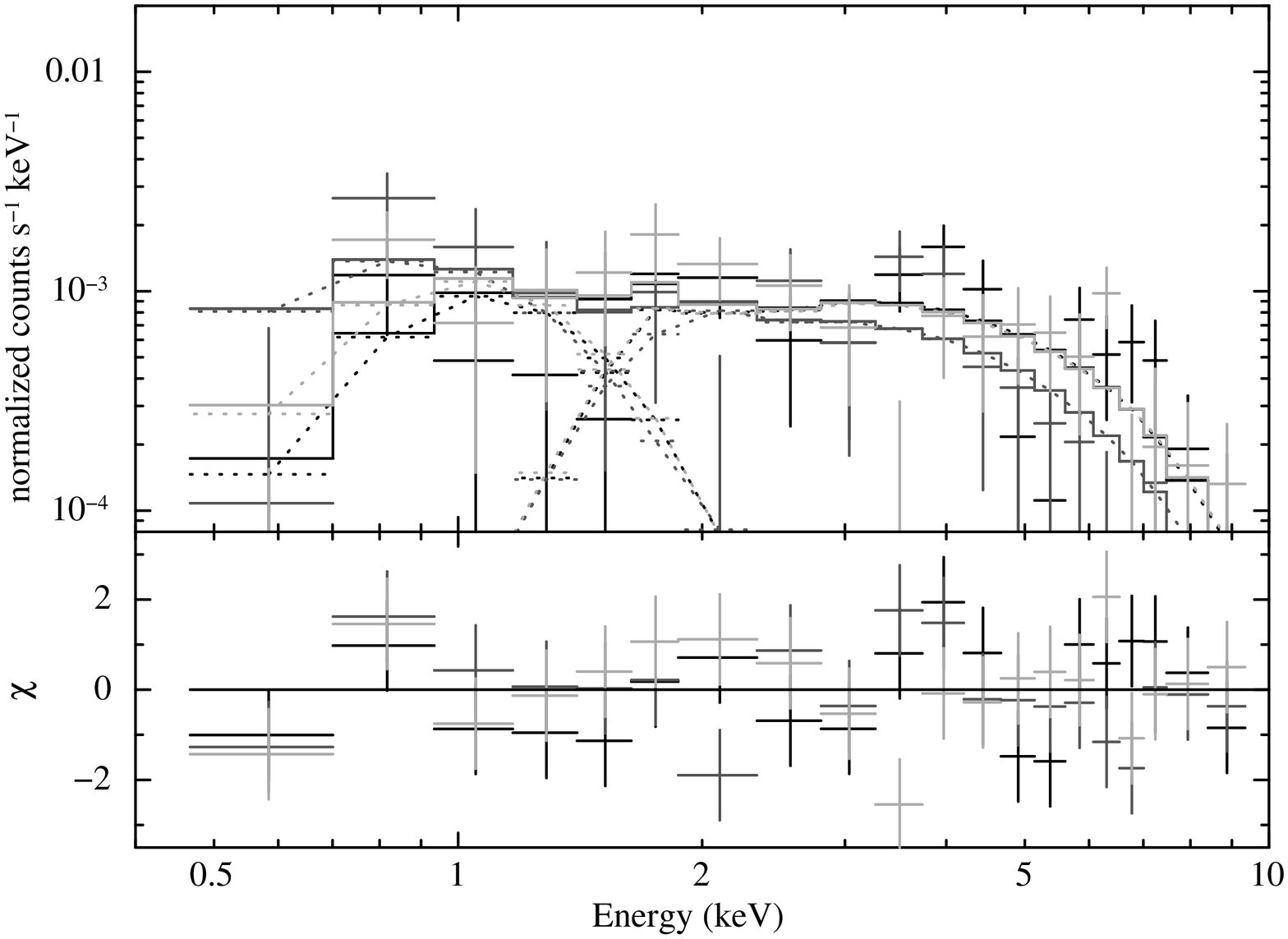}
%    \FigureFile(80mm,80mm){figure6.eps}
  \end{center}
  \caption{
XIS spectra of Src~A. Black, Dark Gray and Light Gray lines
represent the data and model for the XIS0, XIS1 and XIS3,
respectively. The contribution of Chandra point sources is indicated
by the dotted lines in the soft-band. The additional dotted
lines in the hard-band show the diffuse component.
}\label{fig:6th}
\end{figure}

The Suzaku spectra of Src~A in the 0.4--2~keV band were
compared with simulated spectra of the Chandra sources
within the Src~A region.  In the simulation, we took
into account the leakage of photons from Srcs~1, 2, and 3,
and the contamination from Srcs~4 and 5.  Since the
count rates of Srcs~4 and 5 were quite small (table~\ref{tab:1st}), a
qualitative analysis of the combined spectrum of Srcs~4 and 5 was
impossible.  We assume that Srcs~4 and 5 have the same spectral
shape as the best-fit model to the Chandra Srcs~1, 2, and 3.  The
count rates of the Chandra
sources were varied, but the ratios between them were fixed
according to the values in table \ref{tab:1st}.  We found that the Suzaku
spectra in the 0.4--2 keV band can be explained
($\chi^2/d.o.f = 20.37/15$) when the total flux of the Chandra Srcs~1, 2,
and 3 is $(2.6 \pm 1.3) \times 10^{-14}$ {\rm erg cm$^{-2}$ s$^{-1}$} (figure \ref{fig:6th}).
This value is consistent with the above Chandra result of
$(3.4 \pm 1.0) \times 10^{-14}$ {\rm erg cm$^{-2}$ s$^{-1}$},
and it is reasonable to conclude that the Suzaku soft-band spectra
of Src~A can be explained by the Chandra point sources.

On the other hand, the Suzaku spectra of Src~A in the
2--10~keV band cannot be described by the Chandra point
sources alone having soft X-ray emission. In the model
fitting of the Suzaku spectra of Src~A in the 0.4--10~keV
band, we added a power-law component modified by
interstellar absorption to explain the remaining hard X-ray
emission.  A hydrogen column density ($N_{\rm H}$), a photon
index ($\Gamma$), and a normalization were free
parameters. The spectra of Src~A with the best-fit model are
shown in figure \ref{fig:6th}; the best-fit parameters are
$N_{\rm H} = 3.4^{+2.4}_{-1.6} \times 10^{22}$~cm$^{-2}$, 
$\Gamma = 2.0^{+0.9}_{-0.7}$, and $\chi2/d.o.f.$=60.49/55.  
The observed
flux of the power-law component in the 2--10 keV band is
$F$(2--10 keV) = $1.4 \times 10^{-13}$ {\rm erg cm$^{-2}$
  s$^{-1}$}.  The contribution of point sources is indicated
by the dotted lines in the soft-band.  The additional dotted
lines in the hard-band show the power-law model.

\subsubsection{Src~B}

\begin{figure}
  \begin{center}
    \FigureFile(160mm,){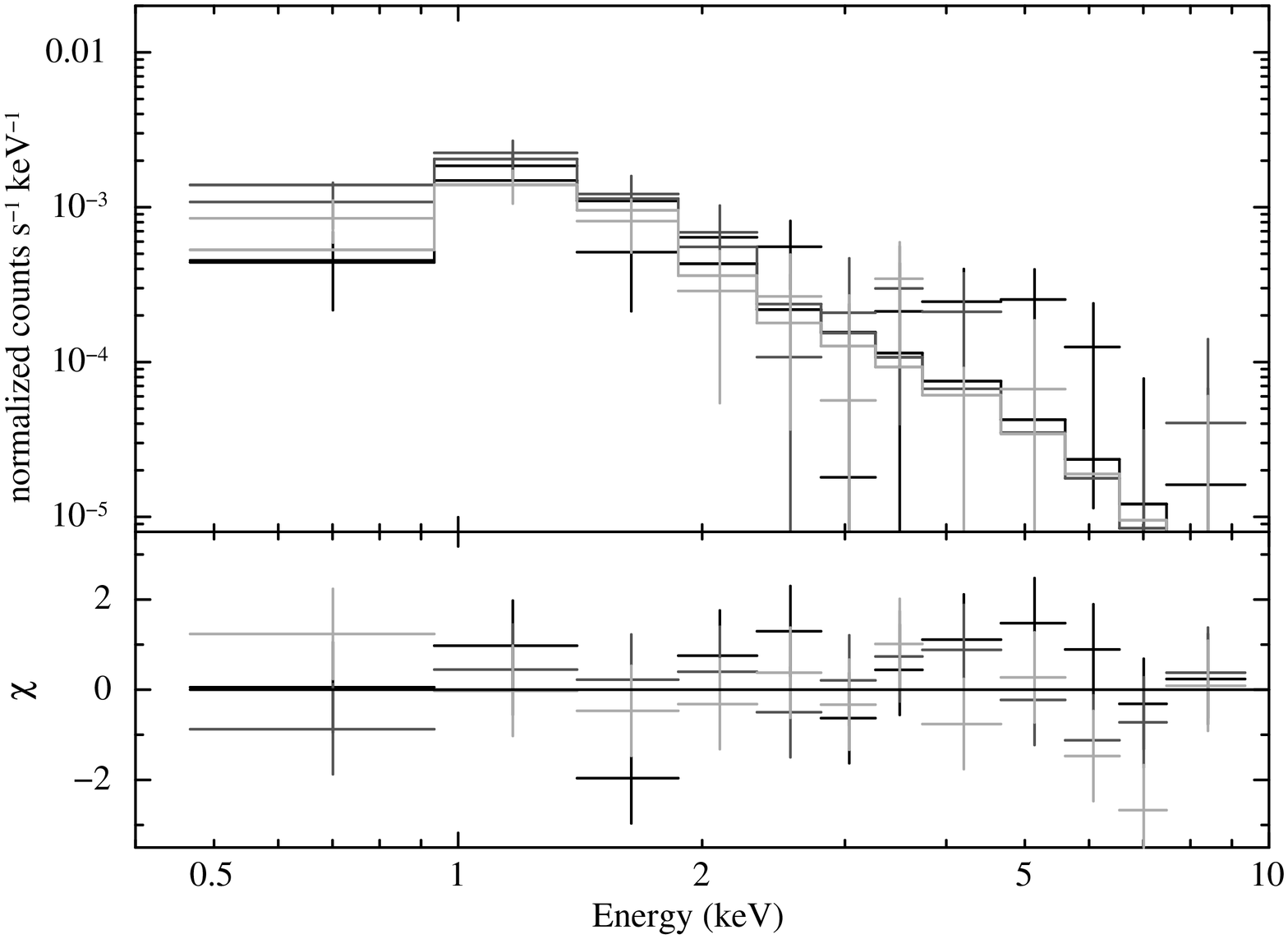}
%    \FigureFile(80mm,80mm){figure7.eps}
  \end{center}
  \caption{
XIS spectra of Src~B. Black, Dark Gray and Light Gray lines
represent the data and model for the XIS0, XIS1 and XIS3,
respectively.
}\label{fig:7th}
\end{figure}

The Suzaku spectra of Src~B were extracted from a circular
region centered on Src~B.  We found no counterpart for Src~B
in literatures.  A circle with a radius of \timeform{1.5'}
is used as the source region and an annulus with an inner
radius of \timeform{4.5'} and an outer radius of
\timeform{7'} is used for the background.  The spectra of
Src~B are shown in figure \ref{fig:7th}.  We fitted the
spectra with an absorbed power-law model.  The best-fit
parameters are $N_{\rm H} = 0.3^{+0.3}_{-0.2} \times
10^{22}$ {\rm cm}$^{-2}$, $\Gamma = 3.0^{+1.4}_{-0.8}$, and $\chi2/d.o.f.$=29.88/31.
The observed flux in the 0.4--10 keV band is $F$(0.4--10
keV) = $6.4 \times 10^{-14}$ {\rm erg cm$^{-2}$ s$^{-1}$}.

We also tried to fit an absorbed thermal plasma model (the
APEC model:~\cite{Smi01}).  The thermal model yields a
hydrogen column of zero ($< 1.0 \times 10^{22}$) {\rm
  cm}$^{-2}$, a temperature of $1.9^{+1.1}_{-1.3}$ keV,
an abundance of $0.26$ ($< 1.03$) solar, and $\chi2/d.o.f.$=28.73/30.  The observed
flux in the 0.4--10 keV band is $F$(0.4--10 keV) = $6.4
\times 10^{-14}$ {\rm erg cm$^{-2}$ s$^{-1}$}.

\subsection{Intensity upper limit of PSR~J1648$-$4611}

Since the X-ray emission of PSR~J1648$-$4611 was not
detected by the point source search for the Chandra image,
we estimated the intensity upper limit using the Chandra
data as follows.  The event number in the 0.4--10 keV band
in a circular region with a radius of \timeform{3''}
centered on the pulsar is 3~counts. We set the source region
larger than the point spread function of Chandra to include
the possible extended emission from a compact
PWN~\citep{Kar12}.  A background was estimated using an
annular region with an inner radius of \timeform{2'} and an
outer radius of \timeform{3'}, which was outside of the
Suzaku Src~A region, and the background event number in the
0.4--10keV band was 1300; the event number normalized to the
source region was 0.65~counts.  The event number in the
source region after subtracting the background was estimated
to be $2.35^{+3.01}_{-2.05}$~counts by a numerical
simulation assuming the Poisson statistics; since the 99\%
confidence range was estimated to be
$2.35^{+4.01}_{-3.02}$~counts and was consistent with zero,
we conservatively think that the X-rays from the pulsar is
not statistically significant.  The upper limit on the
pulsar X-rays in the 0.4--10~keV band at the 90\% confidence level
was thus 5.4~counts. 
The count number was converted to the energy
flux in the 0.4--10~keV band by using
webPIMMS\footnote{$\langle$http://heasarc.gsfc.nasa.gov/Tools/w3pimms.html$\rangle$.}
assuming a power-law model with $\Gamma=2$ and a hydrogen
column of $N_{\rm H} = 2 \times 10^{22}$ {\rm cm}$^{-2}$
where we adopted the total Galactic HI column density
towards PSR~J1648$-$4611~\citep{Dic90}. The upper limit on
the X-ray flux of PSR~J1648$-$4611 in the 0.4--10 keV band
was thus calculated to be $1.0 \times
10^{-14}$~erg~s$^{-1}$~cm$^{-2}$, while that on the
unabsorbed flux was $2.3 \times
10^{-14}$~erg~s$^{-1}$~cm$^{-2}$.

\section{Discussion}

\subsection{X-ray emission from PSR~J1648$-$4611}

The X-ray emission from the pulsar PSR~J1648$-$4611 was not
detected in the Chandra image.  The upper limit on the flux
of PSR~J1648$-$4611 yields the upper limit on the luminosity
of $8.8\;d_{5.7}^{2} \times 10^{31}$ {\rm erg~s$^{-1}$} in
the 0.4--10.0 keV band, where $d_{5.7}$ is the distance
scaled to 5.7~kpc.

\citet{Kar12} also analyzed the same Chandra data and
detected no X-rays from the pulsar.  Their upper limit on
the X-ray count in the 0.4--10~keV band in the circular region
with a radius of \timeform{3''} is, however,
3.1~counts, which we converted from their upper limit on the
X-ray flux in the 0.5--8~keV band ($5.2 \times
10^{-15}$erg~s$^{-1}$~cm$^{-2}$) using their assumed model
(a power-law model of $\Gamma=1.5$ and $N_{\rm H} = 1.2
\times 10^{22}$~cm$^{-2}$). This value is only $\sim$60\% of
our estimate. The main reason for the discrepancy is the
difference of the methods for the upper limit; \citet{Kar12}
estimated the upper limit only from the background event
number, while we obtained it using both the source and
background event numbers. Our value is conservative anyway.

The upper limit on the X-ray luminosity in the 0.1--2.4~keV
band is $8.7\;d_{5.7}^{2} \times 10^{31}$~erg~s$^{-1}$.
Following the relation between the spin-down power and X-ray
luminosity of rotation-powered pulsars~\citep{Bec97}, the
expected X-ray luminosity of PSR~J1648$-$4611 in the
0.1--2.4~keV band would be $L_{\rm X} \sim
10^{32}$~erg~s$^{-1}$.  This value is consistent with our
upper limit.  Since pulsed GeV emission with the rotation
period of PSR~J1648$-$4611 is observed with the Fermi
satellite,\footnote{See
  $\langle$https:\slash\slash{}confluence.slac.stanford.edu\slash{}display\slash{}GLAMCOG\slash{}Public+List+of+LAT-Detected+Gamma-Ray+Pulsars$\rangle$.}
it is natural to expect an X-ray
emission of the pulsar. However, even if PSR~J1648$-$4611
would have X-ray emission with the luminosity expected from
the spin-down luminosity, we would not be able to detect it
with the short exposure time ($\sim$10~ks) because of the
heavy extinction due to the Galactic plane and of the
distance.

\subsection{Diffuse hard X-ray emission from Src~A}

We found the diffuse hard X-ray emission from Src~A in the
Suzaku image.  The peak position of the emission is close to
PSR~J1648$-$4611. A natural interpretation is that the
diffuse hard X-ray emission comes from a PWN around
PSR~J1648$-$4611.  The photon index of the hard X-ray
emission is $2.0^{+0.9}_{-0.7}$.  This value is consistent
with the typical values of PWNe~(\cite{Fle07} and references
therein). The column density $N_{\rm
  H} = 3.4^{+2.4}_{-1.6} \times 10^{22}$~cm$^{-2}$ is also consistent with the
Galactic HI column density towards PSR~J1648$-$4611
($\sim 2\times10^{22}$~cm$^{-2}$; \cite{Dic90}).
Furthermore, a constant component in the GeV
emission is reported~\citep{Abr12}, and the component
also suggests the existence of the PWN.

On the other hand, the luminosity of the diffuse hard X-ray
emission is $7.3\;d_{5.7} \times 10^{32}$~erg~s$^{-1}$ in
the 2--10 keV band.  Following the relation between
spin-down power/characteristic age and X-ray luminosity of
PWNe~\citep{Mat09}, the expected X-ray luminosity of the
associated PWN would be $L_{\rm X} \sim
10^{31}$~erg~s$^{-1}$.  Thus our result is much larger than
the expected luminosity, even when the uncertainty
in the distance estimation of $\sim 30\%$~\citep{Kra03} is taken into account.
Furthermore we obtained the lower limit on the ratio
of the PWN to pulsar luminosities $L_{\rm PWN (0.5-8
  keV)}/L_{\rm PSR (0.5-8 keV)} \geq 16$.  This value is
larger than the average $L_{\rm PWN (0.5-8 keV)}/L_{\rm PSR
  (0.5-8 keV)} \sim 4$ reported by~\citet{Kar07}.
Additionally, the physical size of the diffuse X-ray
emission is $\sim 3.3\;d_{5.7} (\theta/\timeform{2'})
{\rm pc}$, where $\theta$ is the angular radius of Src~A.
This size is an order of magnitude smaller than the expected
value based on the PWN size-age relation reported
by~\citet{Bam10}.  Therefore if the diffuse hard X-ray
emission is the PWN of PSR~J1648$-$4611, it is much brighter
than that expected from the spin-down luminosity of the
pulsar, and the spatial extension is rather compact.

\subsection{Connection to HESS~J1646-458}

If the origin of the TeV emission from HESS~J1646$-$458 is
PSR~J1648$-$4611, an unreasonably high efficiency of
$\epsilon_{\gamma} = L_{\gamma} / \dot E =
96\;d_{5.7}^{2}$\% is required for the production of TeV
flux of HESS~J1646$-$458, which is $5.2 \times 10^{-11}$
{\rm cm$^{-2}$ s$^{-1}$} above 0.2~TeV~\citep{Abr12}.
Additionally, the observed VHE $\gamma$-ray emitting region
has a diameter of about \timeform{2D}, which is a factor of
$\sim 30$ larger than that of the X-ray emitting region of
Src~A and is not consistent with the ratio between the
predicted VHE $\gamma$-ray and X-ray size of PWN ($\sim
6:$~\cite{Aha05}).  These results imply that
HESS~J1646$-$458 seems unlikely to be explained only as a
PWN powered by PSR~J1648$-$4611.  However, it is conceivable
that parts of the VHE $\gamma$-ray emission of
HESS~J1646$-$458 are powered by the pulsar.

\section{Summary}

We observed the pulsar PSR~J1648$-$4611 with Suzaku, and
discovered the diffuse hard X-ray emission around
PSR~J1648$-$4611.  The spatial distribution and the photon
index of the diffuse emission suggest that a PWN exists
around the pulsar.  The luminosity of the diffuse emission,
however, is much larger than that expected from the
spin-down luminosity of the pulsar.  Parts of the VHE
$\gamma$-ray emission of HESS~J1646$-$458 could be caused by
this PWN powered by PSR~J1648$-$4611.

\bigskip

We are grateful to H.~Kunieda, and H.~Mori for their useful
comments. We would like to thank all Suzaku members. MS is
supported by Grant-in-Aid for Japan Society for the
Promotion of Science (JSPS) Fellows, 23-5737, HM is
supported by Grant-in-Aid for Scientific Research (B),
22340046. This work is partially supported by the
Grant-in-Aid for Nagoya University Global COE Program,
``Quest for Fundamental Principles in the Universe: from
Particles to the Solar System and the Cosmos'', from the
Ministry of Education, Culture, Sports, Science and
Technology of Japan.

\end{document}